 \documentclass{elsart}

    \usepackage{latexsym,bm,amsmath,amssymb,amsfonts}
    \usepackage{epsfig,graphics,graphicx}
    \usepackage{makeidx}

\long\def\comment#1{ }

\newcommand{\beq}{\begin{eqnarray}}
\newcommand{\eeq}{\end{eqnarray}}
\newcommand{\be}{\vspace{-.4cm}\begin{eqnarray}}
\newcommand{\ee}{\vspace{-.5cm}\end{eqnarray}}

\newcommand{\BQ}{\begin{equation}}
\newcommand{\EQ}{\end{equation}}
\newcommand{\BQA}{\begin{eqnarray}}
\newcommand{\EQA}{\end{eqnarray}}

\def\simge{\mathrel{%
   \rlap{\raise 0.511ex \hbox{$>$}}{\lower 0.511ex \hbox{$\sim$}}}}
\def\simle{\mathrel{
   \rlap{\raise 0.511ex \hbox{$<$}}{\lower 0.511ex \hbox{$\sim$}}}}

\begin{document}

\begin{frontmatter}

\parbox[]{16.0cm}{ \begin{center}
\title{Soft photon anomaly and gauge/string duality}

\author{Yoshitaka Hatta and Takahiro Ueda }

\address{Graduate School of Pure and Applied Sciences, University
of Tsukuba,\\ Tsukuba, Ibaraki 305-8571, Japan}


\begin{abstract}
Motivated by the recent DELPHI report on the anomalous photon production in $e^+e^-$ annihilation,
we exactly calculate the inclusive cross section of soft photons in the strong coupling  limit of ${\mathcal N}=4$ super Yang--Mills. We find that the energy distribution is that of the Bremsstrahlung, while the angular distribution is spherical. Our result elucidates a new non-perturbative source of soft photons not associated with the final state hadronic Bremsstrahlung.
\end{abstract}

\end{center}}

\end{frontmatter}

\section{Introduction}

Conventional wisdom says  that the production of very low energy photons in hadronic collisions can be entirely understood by the Bremsstrahlung from structureless  charged particles in the initial and final states \cite{Landau:1953um,Low:1958sn}. The argument is simple, but strong---soft photons with energy or transverse momentum less than a few hundred MeV  cannot resolve short distance processes occurring at the QCD scale and beyond.  Historically, however, this simple expectation has rarely been borne out  for very low--$p_T$ ($\lesssim 100$ MeV) photons in modern high energy experiments once the collision energy exceeds several tens of GeV.  In 1984, the WA27 Collaboration at CERN observed a clear excess of  soft photons with $p_T<60$ MeV over the Bremsstrahlung prediction in $K^+p$ collisions at 70 GeV \cite{Chliapnikov:1984ed}. Subsequently, similar observations  were reported in various hadronic reactions such as $\pi^\pm p$ and $pp$  collisions  \cite{Botterweck:1991wf,Banerjee:1992ut,Belogianni:1997rh,Belogianni:2002ib,Belogianni:2002ic}. It was gradually recognized that  the resolution of this puzzle was quite challenging. A number of theoretical ideas have been proposed  trying to pin down the origin of the anomalous photons \cite{Andersson:1988nk,Shuryak:1989vn,Lichard:1990ye,Botz:1994bg,Pisut:1995vu}.
 Some of these models met with partial success, but to date no model has been able to give a consistent description of the events as a whole. The deeper problem is that  generally accepted theories of soft photon production beyond the QED Bremsstrahlung are nonexistent, and in this sense we do not understand the puzzle at present any more than we did back in the 80's.\footnote{The situation is better for hard (`prompt') photons  where perturbative QCD can give a good description of the data in most cases. See, e.g., \cite{Aurenche:2006vj} and references therein.}

The problem recently resurfaced with greater seriousness when the DELPHI Collaboration at CERN presented thorough and  very precise  analyses of  the associated photon production in $e^+e^-$ annihilation into hadrons
 \cite{Abdallah:2005wn,ko}. After subtracting the hadronic decay contributions, the number of soft photons was found to be larger than the theoretical expectation by a factor of about 4. On the other hand, photon yields from $e^+e^-$ annihilation into $\mu^+\mu^-$ pairs do agree with theory \cite{:2007zh}, so the anomaly has uniquely  to do with the strong interaction.  Curiously, however, the {\it shape} of the distribution is consistent with the Bremsstrahlung \cite{Belogianni:2002ib}, namely,
\beq
\frac{dN}{dk} \sim \frac{A}{k^\alpha}\,, \qquad \alpha\approx 1\,, \label{brem}
\eeq
 where $k=|\vec{k}|$ is the photon energy. The data show that the emission from hadrons in the final state contributes to only a fraction of the normalization factor $A$. One then naturally asks whether the Bremsstrahlung, or whatever production mechanisms that lead to the $1/k$ spectrum, in the {\it non-confining} phase can account for the rest.  Implicit in this question is the assumption that hadronization takes a long time  so that, after all, the experimental `soft' photons are not soft enough.  As a matter of fact, this stance has been more or less the common denominator underlying the previous theoretical attempts. [For a recent work, see \cite{Wong:2010gf}.] The difficulty then, of course, is that the problem is clearly non-perturbative, albeit non-confining, so first principle calculations are impractical.

The situation may change, however, with the advent of the AdS/CFT correspondence, or more generally, gauge/string duality \cite{Aharony:1999ti}. At least in a class of non-Abelian gauge theories closely related to QCD, it has now become possible to do exact non-perturbative calculations by investigating the dual string theory in a curved space-time. The best--known example is ${\mathcal N}=4$ supersymmetric Yang--Mills (SYM) theory which is dual to type IIB superstring theory on $AdS_5\times S^5$. Since this theory is non-confining, its strong coupling regime could serve as an insightful model of the pre-hadronization stage in high energy processes. Based on this expectation, in the present work we calculate the inclusive photon cross section in $e^+e^-$ annihilation in strongly coupled ${\mathcal N}=4$ SYM.

  Previously, several authors have  studied hadron production in $e^+e^-$ annihilation from gauge/string duality \cite{Evans:2007sf,BallonBayona:2007rs,Hatta:2008tx,Hatta:2008tn,Hatta:2008qx,Csaki:2008dt,Patino:2009uq,Evans:2009py}. In particular, the energy distribution of hadrons was shown to be  exponential (`thermal')  \cite{Hatta:2008tn,Hatta:2008qx}
\beq
\frac{dN}{dk} \propto e^{-E_k/\Lambda}\,, \label{exp}
\eeq
 where the parameter $\Lambda$ (`temperature') is proportional to the confinement scale of the theory. Because photons do not belong to the particle content of  ${\mathcal N}=4$  SYM (just like photons are external particles to QCD), their distribution is expected to be different from (\ref{exp}). Indeed,  we shall find in Section 3 that the distribution in the $k\to 0$ limit has precisely the form (\ref{brem}) whose coefficient $A$ is exactly calculable. We regard this as a novel source of soft photons not associated with the hadronic Bremsstrahlung, and thereby suggest an interesting avenue toward understanding  the soft photon problem within gauge/string duality.

\section{Preliminaries}
In this section, we first set up our notations and formulate the problem in ${\mathcal N}=4$ SYM. We then collect necessary ingredients for the actual calculation from the type IIB supergravity.

\subsection{Field theory side}

Consider $e^+e^-$ annihilation at the center--of--mass energy $\sqrt{s}=Q$.
The inclusive cross section of  photons with the four--momenta $k^\mu=(k,\vec{k})$  is given by the standard formula\footnote{Our metric convention is $\eta^{\mu\nu}=\mbox{diag}(-1,1,1,1)$.}
\beq
d\sigma &=& \frac{1}{2Q^2}\frac{d^3\vec{k}}{2k(2\pi)^3}\frac{e^4}{Q^4}\frac{1}{4}\sum_{spin}\bar{u}(p)\gamma_{\mu}
v(p')\bar{v}(p')\gamma_{\mu'} u(p) \nonumber \\
&& \qquad \quad  \times \sum_{X,pol} \int d^4x\, e^{-iq\cdot x} \langle 0|J^\mu(x)|kX\rangle \langle kX|J^{\mu'}(0)|0\rangle
\nonumber \\ \quad
&=& \frac{e^6}{2Q^6} \frac{d^3\vec{k}}{2k(2\pi)^3}(p_\mu p'_{\mu'} +p_{\mu'} p'_\mu -\eta_{\mu\mu'}p\cdot p')
\nonumber \\ &&  \ \times \sum_{pol} \int d^4x d^4y d^4 z\, e^{-iq\cdot x+ik\cdot (y-z)} \langle 0|\tilde{\mbox{T}}\{J^\mu(x)\, \varepsilon_k\cdot J(y)\}\, \mbox{T}\{\varepsilon^*_k \cdot J(z) J^{\mu'}(0)\}|0\rangle\,,
\eeq
where  $p^\mu=(Q/2,0,0,Q/2)$,  $p'^\mu=(Q/2,0,0,-Q/2)$ and $q^\mu=(Q,0,0,0)$ are the four--momenta of the electron, the positron and the virtual photon, respectively. $J^\mu$ is the electromagnetic current operator and the sum is over the photon polarizations $\varepsilon_{k}^{i=1,2}$. The symbols $\mbox{T}$ and $\tilde{\mbox{T}}$ denote the time--ordering and anti--time--ordering products, respectively. It is convenient to rewrite the matrix element using the closed--time--path (CTP) formalism  \cite{Keldysh:1964ud,Balitsky:1990ck}
\beq d\sigma &=&  \frac{e^6}{2Q^6} \frac{d^3\vec{k}}{2k(2\pi)^3} (p_\mu p'_{\mu'} +p_{\mu'} p'_\mu -\eta_{\mu\mu'}p\cdot p')
\sum_{pol}  \int d^4x d^4y d^4 z\, e^{-iq\cdot x+ik\cdot (y-z)}
\nonumber \\ &&
\qquad \qquad \qquad \qquad \qquad \times  \langle 0|\mbox{T}_{\mathcal{C}}\{J_{(2)}^\mu(x) \varepsilon_k\cdot J_{(2)}(y) \varepsilon_k^*\cdot J_{(1)}(z) J_{(1)}^{\mu'}(0)\}|0\rangle\,, \label{base}
\eeq
  where  $\mbox{T}_{\mathcal{C}}$ denotes the contour--ordering product. The subscripts (1) and (2) on $J$ indicate that the operator lives on the forward and backward branches of the closed time path, respectively.

 The goal of this paper is to evaluate (\ref{base}) in the soft photon region $k\ll Q$ in strongly coupled ${\mathcal N}=4$ SYM using the AdS/CFT correspondence. In this theory, the analog of the electromagnetic current $J$ in QCD would be a component of the ${\mathcal R}$--current operator $J^a$ ($1\le a \le 15$) associated with a $U(1)$ subgroup of the $SU(4)$ ${\mathcal R}$--symmetry. The component $J^3$ corresponding to the generator\footnote{We employ the standard normalization of the generators $\mbox{tr}(t^at^b)=\frac{1}{2}\delta^{ab}\,$.} $t^3=\mbox{diag}(1/2,-1/2,0,0)$  is composed of two (out of four) species of Weyl fermions and two (out of three) species of complex scalars of ${\mathcal N}=4$ SYM
\beq J_\mu^3= \frac{1}{2}\left(\bar{\psi}_1\bar{\sigma}_\mu \psi_1 - \bar{\psi}_2 \bar{\sigma}_\mu \psi_2\right)+\frac{i}{2}\phi_1^\dagger\left(D_\mu -\overleftarrow{D}_\mu\right) \phi_1 +\frac{i}{2}\phi_2^\dagger\left(D_\mu -\overleftarrow{D}_\mu\right) \phi_2\,. \label{use}  \eeq

It is known that in ${\mathcal N}=4$ SYM  the two-- and three--point functions of the ${\mathcal R}$--current operators are subject to non-renormalization theorems and are therefore independent of the 't Hooft coupling \cite{Freedman:1998tz}. However, this does not apply (to our knowledge) to the four--point functions.
In Appendix A, we calculate the cross section (\ref{base}) in ${\mathcal N}=4$ SYM at vanishing coupling, namely, in the `parton model'. The result is indeed different from the one at strong coupling to be presented in the next section.

Before moving on to the gravity description, a comment is in order regarding the normalization of the cross section. In ${\mathcal N}=4$ SYM, the fields are in the adjoint representation of the color group. Accordingly, (\ref{base}) will be proportional to $N_c^2$, whereas in QCD it is proportional to $N_c$. In order to alleviate this difference, we find it convenient to divide (\ref{base}) by the total cross section $\sigma_{tot}$ of $e^+e^-$ annihilation which also scales as $N_c^2$. Since the latter is proportional to the two--point correlation function of $J$'s, it can be exactly evaluated to be
\beq
\sigma_{tot}=\frac{e^4N_c^2}{32\pi Q^2}\,. \label{tot}
\eeq
Thus we shall present the results in the form of the photon yield
\beq
k\frac{dN}{d^3\vec{k}}&\equiv & \frac{k}{\sigma_{tot}}\frac{d\sigma}{d^3\vec{k}} \nonumber \\
 &=& \frac{e^2}{\pi^2 N_c^2Q^4} (p_\mu p'_{\mu'} +p_{\mu'} p'_\mu -\eta_{\mu\mu'}p\cdot p')  \sum_{pol}  \int d^4x d^4y d^4 z\, e^{-iq\cdot x+ik\cdot (y-z)} \nonumber \\
 &&  \qquad \qquad \qquad \qquad \qquad \times  \langle 0|\mbox{T}_{\mathcal{C}}\{J_{(2)}^\mu(x) \varepsilon_k\cdot J_{(2)}(y) \varepsilon_k^*\cdot J_{(1)}(z) J_{(1)}^{\mu'}(0)\}|0\rangle\,, \label{last}
 \eeq
  which is independent of $N_c$.
Actually, the quantity which directly comes out of the AdS/CFT calculation  is
\beq
  \sum_{pol} \int d^4x d^4y d^4 z\, e^{-iq\cdot x+ik\cdot (y-z)} \langle 0|\mbox{T}_{\mathcal{C}}\{\varepsilon_q^*\cdot J_{(2)}(x) \varepsilon_k\cdot J_{(2)}(y) \varepsilon_k^*\cdot J_{(1)}(z) \varepsilon_q\cdot J_{(1)}(0)\}|0\rangle\,, \label{fac}
 \eeq
 where $\varepsilon_q^\mu$ is the virtual photon polarization vector, and the polarization sum is only over $\varepsilon_k$. So our strategy to compute (\ref{last}) will be  to first evaluate (\ref{fac}), then factor out the polarization vectors $\varepsilon^*_{q\mu}\varepsilon_{q\mu'}$, and finally contract the remainder  with the leptonic tensor.

\subsection{Gravity side}
 In the Poincar\'e coordinates, the metric of $AdS_5$ takes the form
\beq
ds^2=g_{mn}dx^m dx^n=\frac{\eta_{\mu\nu}dx^\mu dx^\nu +dz^2}{z^2}\,,
\eeq
 where we use the notation $x^m=(x^\mu,z)$. According to the AdS/CFT correspondence, the strong 't Hooft coupling limit of ${\mathcal N}=4$ SYM is dual to the type IIB supergravity on $AdS_5$. The part of the supergravity action which will be relevant in this paper involves the graviton $g_{mn}$, the dilaton $\phi$ and the $SO(6)\cong SU(4)$ gauge boson $A_m^a$ ($1\le a \le 15$). The action in the Einstein metric reads
 \beq
 S=\frac{1}{2\kappa^2} \int d^5 x \sqrt{-g} \left(\mathfrak{R} -\frac{4}{3}\partial_m \phi \partial^m \phi
  \right)&-&\frac{1}{4g_{YM}^2}\int d^5 x F^a_{mn}F^{mn}_a e^{-\frac{4}{3}\phi}  \nonumber \\ &+&
\frac{N^2_c}{96\pi^2}\int d^5 x \, \epsilon^{mnpqr} d^{abc} \partial_m A_{n}^a \partial_p A_{q}^b A_r^c\,, \label{act}
  \eeq
  where $2\kappa^2=8\pi^2/N_c^2$ and $g_{YM}^2=(4\pi/N_c)^2$.  The last term is the Chern--Simons term whose normalization has been determined in \cite{Freedman:1998tz}. It is proportional to the totally symmetric d--symbol $d^{abc}=2\mbox{tr}(\{t^a,t^b\}t^c)$ of $SU(4)$.

The current operator $J^3_\mu$ inserted in the matrix element (\ref{fac}) defined on the  boundary Minkowski space excites a component of the $SO(6)$ gauge boson $A^{3}_m$ in the bulk $AdS_5$. This obeys the five--dimensional Maxwell equation supplemented by a gauge condition which we take to be
\beq
\partial_\mu A^\mu +z\partial_z \left(\frac{A_z}{z}\right)=0\,.
\eeq
(We suppress the $SU(4)$ superscripts when unnecessary.)
The solutions are, for the outgoing lightlike photon with $k^\mu k_\mu=0$,
\beq
A_\mu (x^m, k)=A_\mu^*(x^m, -k)=\varepsilon_\mu^* (k)\, e^{-ik\cdot x}\,, \qquad A_z(x^m,k)=0\,, \label{x}
\eeq
and for the incoming timelike photon with $q^\mu q_\mu=-Q^2$,
\beq
A_\mu(x^m, -q)=A_\mu^*(x^m,q)=i\varepsilon_\mu(q)\, e^{iq\cdot x}\frac{\pi Qz}{2}H_1^{(1)}(Qz)\,, \nonumber \\
A_z(x^m,-q)=A_z^*(x^m,q)=-q\cdot \varepsilon_q \, e^{iq\cdot x}\frac{\pi z}{2}H_0^{(1)}(Qz)\,, \label{y}
\eeq
 normalized such that $A_{\mu}(z\to 0)=\varepsilon_\mu(q)\,e^{iq\cdot x}$.
We have kept $\varepsilon_q$ to be generic, while required $\varepsilon_k$ to be the physical polarization $\varepsilon_k\cdot k=0$.
The components of the field strength tensor are
  \beq
&& F_{\mu\nu}(z,k)=F^*_{\mu\nu}(z,-k)=-i\left(k_\mu \varepsilon^*_\nu(k) - k_\nu \varepsilon^*_\mu(k) \right) \,, \nonumber \\
&&  F_{\mu z}(z,k)=0\,, \nonumber \\
 &&F_{\mu\nu}(z,-q)=F^*_{\mu\nu}(z,q)=-\left(q_\mu \varepsilon_\nu(q) - q_\nu \varepsilon_\mu(q)\right) \frac{\pi zQ}{2}H_1^{(1)}(Qz)\,, \nonumber \\
 &&F_{\mu z}(z,-q)=F_{\mu z}^*(z,q)=i\left(\varepsilon_\mu(q)\, q^2-q_\mu \, q\cdot \varepsilon_q\right) \frac{\pi z}{2}H^{(1)}_0(Qz)\,, \label{maxwell}
  \eeq
where we Fourier--transformed to the momentum space in the flat 3+1 directions.

\section{Four--point correlation function}
\begin{figure}
  \begin{center}
    \includegraphics{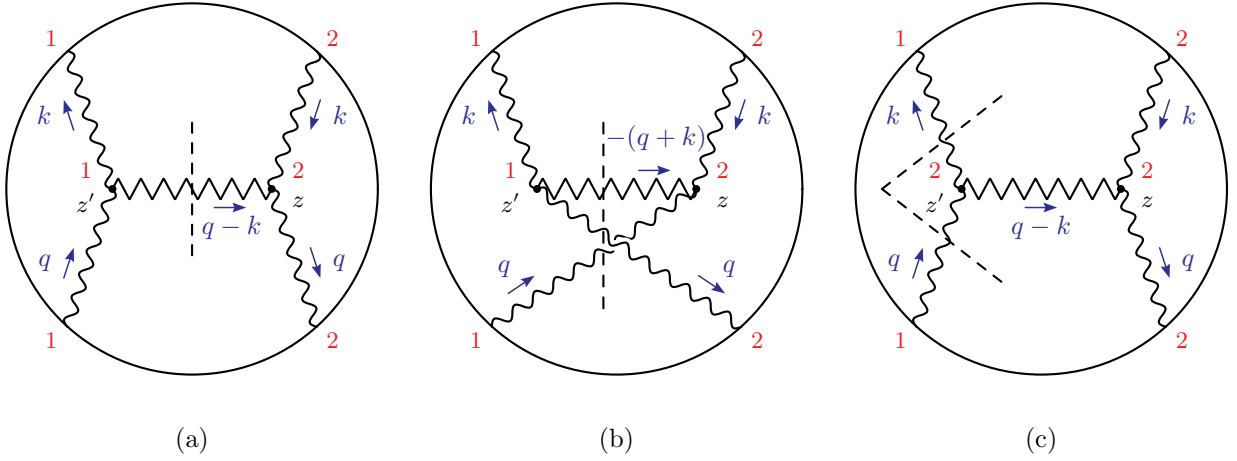}
    \caption{Examples of diagrams for the four--point correlation function (\ref{fac}). The wavy lines represent gauge bosons $A^3_m$. The zigzag line is the  propagator of the graviton, the gauge boson and the dilaton. Propagators which connect vertices on the different time branches are cut by a dashed line. Diagrams (b) and (c) and many others (not shown) do not contribute because of the kinematics.}
    \label{diags}
  \end{center}
\end{figure}

The four--point function of the ${\mathcal R}$--current operator in ${\mathcal N}=4$ SYM has been previously studied both at weak coupling \cite{Bartels:2008zy} and strong coupling \cite{Yoshida:2009dw,Bartels:2009sc,Cornalba:2009ax,Cornalba:2010vk}.
[Real photon production has been studied in \cite{CaronHuot:2006te,Gao:2009se,Marquet:2010sf} in different contexts.]
 At strong coupling, the correlation function is given by the sum of diagrams like shown in Fig.~\ref{diags} where various supergravity modes are exchanged between the bulk $A^3$ fields (\ref{x}) and (\ref{y}). The modes which have a three--point vertex with two $A^3$'s are the graviton, the gauge boson and the dilaton, as can be seen  from the action (\ref{act}).  It turns out that these three modes are equally important for the soft photon production. This is in contrast to the typical high energy limit where one needs to consider only  one component of the graviton propagator $G_{mn,m'n'}=G_{++,--}$ exchanged in the $t$--channel. [The signs $\pm$ refer to the light--cone coordinates.] Our process of interest is considerably more involved  because, as we shall see, it requires {\it all} components of the graviton, the gauge boson and the scalar propagators. Another complication is that, instead of the familiar time--ordered product, (\ref{fac}) features the contour--ordered
   product which involves the doubling of the field degrees of freedom.
  This means that one has to evaluate all possible diagrams in which the three--point vertices are labeled as either  (1) or (2). Fortunately,  the kinematics of the problem allows only the diagrams of the type Fig.~\ref{diags}(a) to be non-vanishing.  We thus focus on this diagram and calculate the three types of exchanges one by one.

\subsection{Graviton exchange}

The graviton exchange contribution to (\ref{fac}) can be evaluated as
\beq
\frac{2\kappa^2}{g_{YM}^4}\int \frac{dz}{z^5}\int \frac{dz'}{z'^5} T^{mn}_{(2)}(z,q,-k)G^{(21)}_{mn;m'n'}(z,z',q-k)T_{(1)}^{m'n'}(z',-q,k)\,. \label{grav}
\eeq
 $G^{(21)}$ is the cut (Wightman) propagator of the graviton connecting vertices on different time branches. In order to obtain this, let us first consider the time--ordered propagator of the graviton $G^{(11)}$ in $AdS_5$ \cite{D'Hoker:1999jc}
  \beq
 G^{(11)}_{mn,m'n'}= \bigl(\partial_m \partial_{m'}u \partial_n \partial_{n'} u + \partial_m \partial_{n'} u \partial_n \partial_{m'} u\bigr) G(u) + g_{mn}g_{m'n'}H(u)+\cdots \,,
 \eeq
  where we omitted gauge artifacts. $u$ is the chordal distance
  \beq
  u=\frac{(z-z')^2+\eta_{\mu\nu}(x-x')^\mu (x-x')^\nu }{2zz'}\,, \label{cho}
  \eeq
   and $H(u)$ is given as \cite{D'Hoker:1999pj}
   \beq
   H(u)=
     -2(1+u)^2G(u)+\frac{4}{3}(1+u)G_3(u)\,.
    \eeq
   In the above, $G$ and $G_3$ are the scalar propagators in $AdS_5$ with mass squared $m^2=0$ and $m^2=-3$, respectively.
    \beq
   &&  \square G=\left(u(u+2)\partial_u^2 +5(u+1)\partial_u \right)G = \left(z^2(\partial_z^2+\partial^2_\mu)-3z\partial_z\right)G \nonumber \\
    && \qquad = i z^5 \delta(z-z')\delta^{(4)}(x-x')\,. \nonumber \\
    && (\square +3)G_3=i z^5\delta(z-z')\delta^{(4)}(x-x')\,.
    \eeq
    For the present purpose, the solutions are most conveniently expressed as
    \beq
    G&=& z^2z'^2 \int \frac{d^4 p}{(2\pi)^4}e^{ip\cdot (x-x')} \int_0^\infty
    d\omega \frac{-i \omega}{\omega^2+p^2-i\epsilon}J_2(\omega z)J_2(\omega z')\,,
    \nonumber \\
     G_3&=& z^2z'^2 \int \frac{d^4 p}{(2\pi)^4}e^{ip\cdot (x-x')} \int_0^\infty
    d\omega \frac{-i \omega}{\omega^2+p^2-i\epsilon}J_1(\omega z)J_1(\omega z')\,.
    \eeq
     Indeed, in this representation one recognizes the familiar four--dimensional Minkowski propagator.
 From this, the cut propagator $G^{(21)}$ can be obtained by a simple replacement \cite{Satoh:2002bc} in $G$ and $G_3$
  \beq
  \frac{-i}{\omega^2+(q-k)^2-i\epsilon} &\to& 2\pi \theta(q^0-k^0) \delta(\omega^2 +(q-k)^2)\,. \label{pole}
  \eeq
    The presence of the theta function in cut propagators  guarantees that, for instance, the diagram (b) of Fig.~\ref{diags} vanishes because the intermediate energy is negative $-(q^0+k^0)<0$. Similarly, cutting the bulk--to--boundary propagators gives the theta functions $\theta(\pm q^0)$ and $\theta(\pm k^0)$. Using these constraints, one can check that all the diagrams except for Fig.~\ref{diags}(a) vanish.

 Returning to (\ref{grav}), the `energy momentum tensor' is given by
  \beq
  T^{mn}_{(1)}(-q,k)&=&(T_{(2)}^{mn}(q,-k))^*
  \nonumber
 \\ &=&\frac{1}{2}\left(F^{m}_{\ \ l}(-q) F^{nl}(k) + F^{n}_{\ \ l}(-q) F^{ml}(k) \right)  -\frac{g^{mn}}{4}F_{pq}(-q)F^{pq}(k)\,,
  \eeq
  where  $F_{mn}$ is as in (\ref{maxwell}).
 Explicitly,
 \beq
  T^{\mu\nu}(z,-q,k) &=&
  iz^7\frac{\pi Q}{2}H_1^{(1)}(Qz)\Bigl(\varepsilon_q\cdot \varepsilon^*_k \, q^\mu k^\nu - q\cdot \varepsilon^*_k \, \varepsilon_q^{\mu} k^\nu - k\cdot \varepsilon_q \,  q^\mu\varepsilon_k^{*\nu}  + q\cdot k \, \varepsilon_q^{\mu}\varepsilon_k^{*\nu}
    \nonumber \\ && \qquad \qquad \qquad \qquad \qquad -\frac{\eta^{\mu\nu}}{2}\left(\varepsilon_q\cdot \varepsilon^*_k \, q\cdot k-q\cdot \varepsilon^*_{k}\, k\cdot \varepsilon_q\right)\Bigr)
    \nonumber \\
   &\equiv & iz^7\frac{\pi Q}{2}H_1^{(1)}(Qz)\left(A^{\mu\nu}-\frac{\eta^{\mu\nu}}{4}A^\rho_{\ \rho}\right)\,. \label{tens}
  \eeq
  (Symmetrization in $\mu\leftrightarrow \nu$ is understood.)
\beq
 T^{zz}(z,-q,k)&=&
  \frac{-iz^7}{2} \frac{\pi Q}{2}H_1^{(1)}(Qz) \left(\varepsilon_q\cdot \varepsilon^*_k \, q\cdot k-q\cdot \varepsilon^*_k\, k\cdot \varepsilon_q \right)\,, \nonumber \\
  T^{\mu z}(z,-q,k)&=&T^{z\mu}(z,-q,k) \nonumber \\
   &=&\frac{z^7}{2}\frac{\pi }{2}H_0^{(1)}(Qz)
   \bigl(-\varepsilon_q\cdot \varepsilon^*_k  \, q^2k^\mu+k\cdot \varepsilon_q \,q^2\varepsilon^{*\mu}_k \nonumber \\
   && \qquad \qquad  \qquad \qquad \quad +q\cdot \varepsilon_k^* \, q\cdot \varepsilon_q k^\mu
   -q\cdot k \, q\cdot \varepsilon_q \,\varepsilon_k^{*\mu}\bigr)\,.
  \eeq

 We now decompose  the full amplitude as
\beq
\int_{z,z'}T^{mn}G_{mn,m'n'}T^{m'n'}&=&\int_{z,z'}T^{\mu\nu}G_{\mu\nu,\mu'\nu'}T^{\mu'\nu'}
\nonumber \\ &+&\int_{z,z'} \left(T^{\mu\nu}G_{\mu\nu,z'z'}T^{z'z'}
+ T^{zz}G_{zz,\mu'\nu'}T^{\mu'\nu'}
+ T^{zz}G_{zz,z'z'}T^{z'z'} \right) \nonumber \\
&+& 4\int_{z,z'}T^{\mu z}G_{\mu z,\mu' z}T^{\mu' z}\nonumber \\
&+& 2\int_{z,z'} \bigl(T^{\mu\nu}G_{\mu\nu,\mu'z'}T^{\mu'z'} + T^{\mu z}G_{\mu z,\mu'\nu'}T^{\mu'\nu'} \nonumber \\
&&  \qquad \qquad \quad + T^{\mu z}G_{\mu z,z'z'}T^{z'z'}+T^{zz}G_{zz,\mu'z'}T^{\mu'z'} \bigr)
\nonumber \\
&\equiv & I_1+(I_2+I_3+I_4)+I_5+(I_6+I_7+I_8+I_9)\,,
\eeq
where we abbreviated as $\int_{z,z'}\equiv \sum_{pol} \int \frac{dz}{z^5}\int \frac{dz'}{z'^5}$ and regrouped terms in a way that slightly facilitates the following analysis.
Consider $I_1$ first. From now on Lorentz indices are raised and lowered with respect to the Minkowski metric $\eta^{\mu\nu}$. The propagator is
 \beq
 G_{\mu\nu;\mu'\nu'}= \frac{\eta_{\mu\mu'}\eta_{\nu\nu'}+\eta_{\mu\nu'}\eta_{\nu\mu'}}
 {z^2z'^2}G^{(21)}(z,z',q-k)+\frac{\eta_{\mu\nu}\eta_{\mu'\nu'}}{z^2z'^2}H^{(21)}(z,z',q-k) \label{h}
 \eeq
 where
 \beq
 G^{(21)}(z,z',q-k)=\pi z^2z'^2 \int_0^\infty d\omega^2 \delta(\omega^2+(q-k)^2) J_2(\omega z)J_2(\omega z')\,.
 \eeq

The $H$ term in (\ref{h}) does not contribute because $\eta_{\mu\nu}T^{\mu\nu}=0$.
The tensor part becomes, using (\ref{tens})
\beq
&& \sum_{pol} \left(A^{\mu\nu}-\frac{\eta^{\mu\nu}}{4}A^\rho_{\ \rho}\right)^*(\eta_{\mu\mu'}\eta_{\nu\nu'}+\eta_{\mu\nu'}\eta_{\nu\mu'})
\left(A^{\mu'\nu'}-\frac{\eta^{\mu'\nu'}}{4}A^{\rho'}_{\ \rho'}\right)
\nonumber \\ && = \sum_{pol} \left((A^{\mu\nu}+A^{\nu\mu})^*A_{\mu\nu} -\frac{1}{2}A_{\mu}^{*\ \mu}A^{\nu}_{\nu} \right)
\nonumber \\
&&
=2\varepsilon^*_{q\mu}\varepsilon_{q\mu'} \Biggl( (q\cdot k)^2 \left(\eta^{\mu\mu'}-\frac{q^\mu q^{\mu'}}{q^2} \right)
+q^2 \left(k^\mu-\frac{k\cdot q}{q^2}q^\mu \right)\left(k^{\mu'}-\frac{k\cdot q}{q^2}q^{\mu'}\right) \Biggr)\,.
\eeq
 where the summation over the photon polarizations $\varepsilon_k^{(i=1,2)}$ has been done according to the identity
 \beq
 \sum_{i}^{1,2} \varepsilon^\mu_i(k) \varepsilon^{*\mu'}_i(k) = \eta^{\mu\mu'}-\frac{k^\mu n^{\mu'}+k^{\mu'} n^\mu}{k\cdot n}
 - \frac{k^\mu k^{\mu'}}{(n\cdot k)^2}\,,
 \eeq
 where $n^\mu\equiv (1,0,0,0)$. In practice, the following simpler substitutions suffice
 \beq
  \sum_{i}^{1,2} \varepsilon^\mu_i(k) \varepsilon^{*\mu'}_i(k) \to \eta^{\mu\mu'}\,, \qquad  \sum_{i}^{1,2} \varepsilon_i(k) \cdot \varepsilon_i^*(k)\to 2\,. \label{sum}
  \eeq

  Integration over $z$ and $z'$ can be done
  \beq
  \frac{\pi Q}{2}\int_0^\infty dz z^2 J_2(\omega z)H_1^{(1)}(Qz) = \frac{-2i\omega^2}{(Q^2-\omega^2)^2}\,, \label{int}
\eeq
 where we assumed $Q>\omega$, consistently with the delta function constraint $\omega^2=Q^2-2Qk$. We also have to require that $Q\to \sqrt{Q^2+i\epsilon}$ has a positive imaginary part in order for (\ref{int}) to be well--defined. This is indeed the case  because  the Hankel function of the first kind comes from the bulk--to--boundary Feynman propagator $G^{(11)}$. We shall encounter integrals similar to (\ref{int}) frequently in the following, so we collected the relevant formulae in Appendix B. The $\omega$ integral  gives
\beq
\pi \int_0^Q d\omega^2  \delta(\omega^2+ (q-k)^2)\frac{4\omega^4}{(Q^2-\omega^2)^4}
=\frac{\pi}{4k^4}\left(1-\frac{2k}{Q}\right)^2 \approx \frac{\pi}{4k^4}\left(1-\frac{4k}{Q}\right)\,, \label{k4}
\eeq
 where, for the sake of later discussion, we keep the leading and the next--to--leading order terms in the expansion in powers of $k/Q$.
 We thus get
 \beq
 I_1 &=& \frac{\pi}{k^4}\varepsilon_{q\mu}^*\varepsilon_{q\mu'}\left(1-\frac{4k}{Q}\right) \Biggl( \frac{(q\cdot k)^2}{2} \left(\eta^{\mu\mu'}-\frac{q^\mu q^\nu}{q^2} \right)
\nonumber \\
 && \qquad \qquad \qquad \qquad \qquad \qquad  +\frac{q^2}{2} \left(k^\mu-\frac{k\cdot q}{q^2}q^\mu \right)\left(k^{\mu'}-\frac{k\cdot q}{q^2}q^{\mu'}\right) \Biggr)\,.
 \eeq
As explained at the end of Section 2.1, we now remove the factor $\varepsilon_q^{*\mu}\varepsilon_q^{\mu'}$ and contract  with the leptonic tensor
\beq
&& (p_\mu p'_{\mu'} +p_{\mu'} p'_\mu -\eta_{\mu\mu'}p\cdot p')
\nonumber \\ && \qquad  \times \Biggl( \frac{(q\cdot k)^2}{2} \left(\eta^{\mu\mu'}-\frac{q^\mu q^{\mu'}}{q^2} \right)
 +\frac{q^2}{2} \left(k^\mu-\frac{k\cdot q}{q^2} q^\mu \right)\left(k^{\mu'}-\frac{k\cdot q}{q^2}q^{\mu'}\right) \Biggr) \nonumber  \\
&&= \frac{Q^4k^2}{4}(1+\cos^2\theta)\,.
\eeq
In this way we find the contribution from this particular tensor component of the graviton exchange
\beq
k\frac{dN_{G1}}{d^3\vec{k}}
  &=&\frac{e^2}{\pi^2 N_c^2Q^4} \frac{2\kappa^2}{g_{YM}^4}\frac{\pi }{k^4}\left(1-\frac{4k}{Q}\right) \frac{Q^4k^2}{4}(1+\cos^2\theta) \nonumber \\  &=&\frac{\alpha_{em}}{32\pi^2 k^2} \left(1-\frac{4k}{Q}\right)(1+\cos^2\theta)\,.
\eeq
Note that the leading term has the same $k$ dependence as the Bremsstrahlung spectrum (\ref{brem}).

Next, let us consider $I_{2,3,4}$. After some cancelations, the sum becomes, in the coordinate space,
\beq
I_2+I_3+I_4&=&-\int_{z,z'} \frac{2}{z^2z'^2}F^{\mu}_{\ \ \lambda} F^{\nu\lambda}(x-x')_\mu(x-x')_\nu G^{(21)}\frac{F'^2}{4} + (c.c.)
\nonumber \\
 &&  \quad +\int_{z,z'} \frac{F^2}{4}\frac{4(1+u)}{3}G_3^{(21)} \frac{F'^2}{4}\,. \label{ii}
 \eeq

Upon Fourier transforming to the momentum space, $x-x'$ may be replaced by a derivative acting on the delta function in $G^{(21)}$ and $G^{(21)}_3$. [Note that the chordal distance $u$ in the second line contains $(x-x')^2$, see (\ref{cho}).] An inspection of the flow of momenta instructs us to replace
\beq
(x-x')_\mu \to +i\frac{\partial}{\partial q^\mu}\,.
\eeq
[The sign is not important here, but it will be important in later calculations.]
We then convert the $q^\mu$ derivative into a $\omega^2$ derivative and do partial integrations. In effect, this amounts to replacing
\beq
&& \int d\omega^2 \, (x-x')_\mu (x-x')_{\nu} \delta(\omega^2+(q-k)^2)\cdots \nonumber \\
 &&  \qquad \qquad \to \int d\omega^2 \left(-\frac{\partial}{\partial q^{\mu}}\frac{\partial}{\partial q^{\nu}}  \delta(\omega^2+(q-k)^2) \right)\cdots \nonumber \\
 &&  \qquad \qquad \to 2\eta_{\mu\nu}\left[\frac{\partial}{\partial \omega^2}\cdots \right]_{\omega^2
=Q^2-2Qk}-4(q-k)_\mu (q-k)_{\nu} \left[\frac{\partial^2}{\partial^2 \omega^2}\cdots \right]_{\omega^2=Q^2-2Qk}\,.  \label{rep}
\eeq
For a technical reason, in the rest of this subsection we assume that $\varepsilon_q\cdot q=0$. This considerably simplifies the manipulation below. The price to pay however is that the gauge invariance of the result is not manifest. To remedy this, we have independently performed the calculation with $\varepsilon_q\cdot q \neq 0$ using the algebraic calculation system
\texttt{FORM}~\cite{Vermaseren:2000nd}, and confirmed that the following results are gauge invariant.

 After straightforward, but tedious calculations using (\ref{sum}), (\ref{rep}), (\ref{a1}) and (\ref{a2}), the first line on the right hand side of (\ref{ii}) becomes
\beq
2\frac{\pi}{k^4}\left(-\frac{1}{4} -\frac{3}{2}\ln \frac{2k}{Q}+\frac{k}{Q}\left(\frac{3}{4}+ \frac{1}{2}\ln \frac{2k}{Q}\right)\right) \left((k\cdot q)^2\varepsilon_q\cdot \varepsilon_q^* +q^2 k\cdot \varepsilon_q k \cdot \varepsilon_q^*\right)\,,
\eeq
 where the factor of 2 in front is the contribution from the complex conjugate. Here again, we have kept the leading and the next--to--leading terms in the $k\to 0$ limit.   Note the appearance of a logarithmic factor. This is reminiscent of what happens in weak coupling calculations in Appendix A, but unlike in the latter case, here the photon energy $k$ (instead of the parton mass $m$ which is zero) makes the logarithm finite.
Similarly, the second line of (\ref{ii}) becomes, using (\ref{a3}) and (\ref{a4}),
\beq
\frac{\pi}{12k^4}\left((k\cdot q)^2\,\varepsilon_q\cdot \varepsilon_q^* +q^2\, k\cdot \varepsilon_q k \cdot \varepsilon_q^*\right)\,.
\eeq
   ${\mathcal O}(k/Q)$ terms accidentally vanishes for this contribution.

We now move on to the next term
\beq
I_5=\int_{z,z'} \frac{4}{z^2z'^2}T^{\mu z}\left(\eta_{\mu\nu}\left(\frac{z}{z'}+\frac{z'}{z}-1-u\right)-
\frac{(x-x')_\mu (x-x')_{\mu'}}{zz'}\right)G^{(21)} T^{\mu' z}\,.
\eeq
After summing  over $\varepsilon_k^{i=1,2}$, it takes the form
\beq
 I_5&=&Q^2\int dz dz' \frac{\pi Q}{2}H_0^{(2)}(Qz)\frac{\pi Q}{2}H_0^{(1)}(Qz') \Biggl\{k\cdot \varepsilon_q^* \, k\cdot \varepsilon_q \left(\frac{z}{z'}+\frac{z'}{z}\right) \nonumber \\
&-&\frac{1}{zz'}(x-x')_\mu (x-x')_{\mu'}\left(\varepsilon_q^* \cdot \varepsilon_q k^\mu k^{\mu'}-k\cdot \varepsilon_q^* \varepsilon_q^{\mu}k^{\mu'} -k\cdot \varepsilon_q \varepsilon^{*\mu'}_q k^\mu+2k\cdot \varepsilon_q^* k\cdot \varepsilon_q \eta^{\mu\mu'}\right)\Biggr\}G^{(21)}\,. \nonumber
\eeq
Integrating over $z,z',\omega$ using (\ref{rep}), (\ref{a4}) and (\ref{a5}), we obtain
\beq
I_5&=& \frac{\pi}{k^4}\Biggl\{ \left(\frac{3}{2}+\frac{k}{Q}\left(-3+2\ln \frac{2k}{Q}\right)\right) (q\cdot k)^2\varepsilon_q\cdot \varepsilon_q^* \, \nonumber \\
 &&  \qquad \qquad  \qquad +\left(4+\frac{k}{Q}\left(-8+6\ln \frac{2k}{Q}\right)\right)q^2\, k\cdot \varepsilon_q^* k\cdot \varepsilon_q\Biggr\}\,.
\eeq

Finally, $I_6+I_7+I_8+I_9$ becomes
\beq
&& I_6+I_7+I_8+I_9=iQ\int\frac{dz}{z} \int dz' \frac{\pi Q}{2}H_0^{(2)}(Qz)\frac{\pi Q}{2} H_1^{(1)}(Qz')\nonumber \\
 && \qquad \times \Biggl\{ 2k\cdot \varepsilon_q^* \, k\cdot \varepsilon_q \, q^\mu -2 k\cdot \varepsilon_q^* \, q\cdot k \, \varepsilon_q^\mu + \frac{z}{z'}\left(\frac{z}{z'}+\frac{z'}{z} -1-u\right) \nonumber \\
 &&  \qquad \quad \times \left(\varepsilon_q^*\cdot \varepsilon_q \, q\cdot k \,k^\mu -k\cdot \varepsilon_q^* \, q\cdot k \, \varepsilon_q^\mu + k\cdot \varepsilon_q \, k\cdot \varepsilon^*_q \,q^\mu \right)  \Biggr\} (x'-x)_\mu G + (c.c.)\,.
\eeq
This is the most formidable term which, in the momentum space, involves the third derivative of the delta function.
After very tedious calculations using (\ref{a1}), (\ref{a2}), (\ref{a4}) and (\ref{a5}), and integrating by parts three times, we arrive at
\beq
I_6+I_7+I_8+I_9& =&2\frac{\pi}{k^4} \Biggl\{ \left(-\frac{1}{2}+\frac{3}{2}\ln \frac{2k}{Q} +\frac{k}{Q}
\left(\frac{3}{4}-\frac{3}{2}\ln \frac{2k}{Q}\right) \right)  (q\cdot k)^2 \, \varepsilon_q\cdot \varepsilon_q^* \nonumber \\ && \quad  + \left(-2 +\frac{3}{2}\ln \frac{2k}{Q} + \frac{k}{Q}\left(\frac{15}{4}-\frac{7}{2}\ln \frac{2k}{Q} \right) \right) q^2\, k\cdot \varepsilon_q \,k\cdot \varepsilon_q^* \Biggr\}\,.
\eeq

Assembling all the contributions, we obtain
\beq
 \sum_{i=1}^9 I_i=\frac{\pi}{k^4} \varepsilon^*_{q\mu} \varepsilon_{q\mu'} \left\{ \left(\frac{7}{12}-\frac{2k}{Q}\right) (q\cdot k)^2\, \eta^{\mu\mu'} +\left(\frac{1}{12}-\frac{k}{Q}\right) q^2 \,k^\mu k^{\mu'} \right\}\,.
\eeq
Note that the logarithms have disappeared both in the leading and the next--to--leading terms.
   Contraction with the leptonic tensor gives
\beq
&& (p_\mu p'_{\mu'} +p_{\mu'} p'_\mu -\eta_{\mu\mu'}p\cdot p')\left\{ \left(\frac{7}{12}-\frac{2k}{Q}\right) (q\cdot k)^2\, \eta^{\mu\mu'}  +\left(\frac{1}{12} -\frac{k}{Q}\right) q^2 \, k^\mu k^{\mu'} \right\} \nonumber \\
&& \qquad \qquad = Q^4k^2\left\{ \left(\frac{7}{24}-\frac{k}{Q}\right) (1+\cos^2\theta)+\left(\frac{1}{4}-\frac{k}{2Q}\right) (1-\cos^2\theta)\right\}\,.
\eeq
We finally arrive at the total contribution from the graviton exchange diagram
\beq
k\frac{dN_G}{d^3\vec{k}}
 &=& \frac{e^2}{\pi^2 N_c^2Q^4} \frac{2\kappa^2}{g_{YM}^4}  \frac{\pi }{k^4} Q^4k^2 \left\{ \left(\frac{7}{24}-\frac{k}{Q}\right) (1+\cos^2\theta)+\left(\frac{1}{4}-\frac{k}{2Q}\right) (1-\cos^2\theta)\right\} \nonumber \\
 &=& \frac{\alpha_{em}}{16\pi^2 k^2} \left\{ \left(\frac{7}{12}-\frac{2k}{Q}\right) (1+\cos^2\theta)+\left(\frac{1}{2}-\frac{k}{Q}\right) (1-\cos^2\theta) \right\}\,. \label{g}
\eeq

\subsection{Gauge boson exchange}
Next we consider the gauge boson exchange. The ordinary three--point coupling from the Yang--Mills action does not contribute because it is proportional to $f^{abc}$ and we have $a=b=3$ in the external states. However, the anomalous coupling from the Chern--Simons term gives a nonvanishing contribution
\beq
&&g_{YM}^2 \left(\frac{N_c^2 }{64\pi^2}\right)^2 d^{33c}d^{33c}\int dz dz' \epsilon^{mnpqr} \epsilon^{m'n'p'q'r'} \nonumber \\
&& \qquad \qquad \qquad  \qquad \times F_{mn}(q)F_{pq}(-k)G^{(21)}_{rr'}(z,z',q-k) F_{m'n'}(-q)
F_{p'q'}(k)\,, \label{gauge}
\eeq
 where the gauge boson propagator in the coordinate space is \cite{D'Hoker:1999jc} \beq
 G_{rr'}=-\partial_r \partial_{r'} u\, G_3(u)+\cdots\,,
 \eeq
  up to gauge artifacts. The cut propagator $G^{(21)}_{rr'}$ can be obtained in the same way as before. The gauge boson exchange from the Chern--Simons coupling was previously considered in \cite{Hatta:2009ra,Bartels:2009sc} in the high energy limit where only one component of the propagator $G_{rr'} = G_{+-}$ was necessary. Here, however, all the components of $G_{rr'}$ will be important.
   The group factor  in (\ref{gauge}) is, using $t^3=\mbox{diag}(1/2,-1/2,0,0)$,
 \beq
 d^{33c}d^{33c}=\frac{1}{2}\,.
 \eeq
 Note that  $d^{333}=0$, so the intermediate gauge boson $A^c$ must carry a different $SU(4)$ index $c\neq 3$.

As before, we decompose the amplitude as
\beq
&& \int_{z,z'} \epsilon^{mnpqr} \epsilon^{m'n'p'q'r'} F_{mn}(q)F_{pq}(-k)G_{rr'} F_{m'n'}(-q)F_{p'q'}(k) \nonumber \\
 && \qquad =\int_{z,z'} \epsilon^{\mu\nu\rho\lambda} \epsilon^{\mu'\nu' \rho' \lambda'}F_{\mu\nu}(q)F_{\rho\lambda}(-k) (-\partial_z\partial_{z'}u) G_3 F_{\mu'\nu'}(-q)F_{\rho'\lambda'}(k)
\nonumber \\  &&
\qquad +4\int_{z,z'} \epsilon^{\mu\nu\rho\lambda}\epsilon^{\mu'\nu'\rho'\lambda'}F_{\mu z}(q)F_{\rho\lambda}(-k)\left(-\partial_\nu \partial_{\nu'}u \right)G_3 F_{\mu'z'}(-q)F_{\rho'\lambda'}(k) \nonumber \\
&&  \qquad -2\int_{z,z'} \epsilon^{\mu\nu\rho\lambda}\epsilon^{\mu'\nu'\rho'\lambda'}F_{\mu z}(q)F_{\rho\lambda}(-k)\left(-\partial_\nu \partial_{z'} u \right) G_3 F_{\mu'\nu'}(-q)F_{\rho'\lambda'}(k) + (c.c.)
\nonumber \\
&& \qquad  \equiv J_1 +J_2+J_3\,,
\eeq
 where this time $\int_{z,z'}\equiv \sum_{pol} \int dz \int dz'$.
The integrals over $z$, $z'$ and $\omega^2$ are simpler than in the graviton case, while the tensor part is quite complicated. We have used  \texttt{FORM} for the latter and obtained
\beq
 J_1&=&-\frac{20\pi}{k^4}\left(1-\frac{8k}{5Q}\right)\varepsilon_q^{*\mu}\varepsilon_q^\nu \Biggl( (q\cdot k)^2 \left(\eta_{\mu\nu}-\frac{q_\mu q_\nu}{q^2} \right)
+q^2 \left(k_\mu-\frac{k\cdot q}{q^2}q_\mu \right)\left(k_\nu-\frac{k\cdot q}{q^2}q_\nu\right) \Biggr)\,,
\nonumber \\
 J_2&=&-\frac{8\pi}{ k^4}\left(1-\frac{2k}{Q}\right) \varepsilon_q^{*\mu}\varepsilon_q^\nu q^2\left(k_\mu-\frac{k\cdot q}{q^2}q_\mu \right)\left(k_\nu-\frac{k\cdot q}{q^2}q_\nu\right)\,,
\nonumber \\
J_3&=& \frac{24\pi }{k^4} \left(1-\frac{4k}{3Q}\right)\varepsilon_q^{*\mu}\varepsilon_q^\nu \Biggl( (q\cdot k)^2 \left(\eta_{\mu\nu}-\frac{q_\mu q_\nu}{q^2} \right)
+q^2 \left(k_\mu-\frac{k\cdot q}{q^2}q_\mu \right)\left(k_\nu-\frac{k\cdot q}{q^2}q_\nu\right) \Biggr)\,. \nonumber \\
\eeq


Summing the three contributions, we get the total gauge boson contribution
\beq
k\frac{dN_A}{d^3\vec{k}}
 &=& \frac{e^2}{\pi^2 N_c^2Q^4} \frac{g_{YM}^2}{2}\left(\frac{N_c^2}{64\pi^2}\right)^2 \frac{4\pi }{k^4}
  (p_\mu p'_\nu +p_\nu p'_\mu -\eta_{\mu\nu}p\cdot p') \nonumber \\ && \times \left\{ (q\cdot k)^2 \left(\eta_{\mu\nu}-\frac{q_\mu q_\nu}{q^2} \right)
-q^2 \left(1-\frac{4k}{Q}\right) \left(k_\mu-\frac{k\cdot q}{q^2}q_\mu \right)\left(k_\nu-\frac{k\cdot q}{q^2}q_\nu\right) \right\}
\nonumber \\
&=& \frac{\alpha_{em}}{16\pi^2k^2}\left\{ \frac{1}{4}(1+\cos^2\theta)+\left(\frac{1}{2}-\frac{k}{Q}\right)(1-\cos^2\theta) \right\}\,. \label{boson}
\eeq

\subsection{Dilaton exchange}
Finally, the dilaton exchange amplitude following from the action (\ref{act}) is
\beq
\frac{3}{8}2\kappa^2 \left(\frac{4}{3}\frac{1}{g_{YM}^2}\right)^2 \sum_{pol} \int \frac{dz}{z^5}\int \frac{dz'}{z'^5} \frac{F^2_{(2)}(q,-k)}{2} G^{(21)} \frac{F^2_{(1)}(-q,k)}{2}\,.
\eeq
By now it is easy to evaluate this. The result is
\beq
&& \sum_{pol}\int \frac{dz}{z^5}\int \frac{dz'}{z'^5} \frac{F^2_{(2)}(q,-k)}{2} G^{(21)} \frac{F^2_{(1)}(-q,k)}{2}
\nonumber \\
&&= \frac{\pi}{4k^4}\left(1-\frac{4k}{Q}\right)\varepsilon^\mu_q \varepsilon_q^{*\nu} \nonumber \\
 &&\qquad  \qquad  \times \left\{ (q\cdot k)^2 \left(\eta_{\mu\nu}-\frac{q_\mu q_\nu}{q^2} \right)
+q^2 \left(k_\mu-\frac{k\cdot q}{q^2}q_\mu \right)\left(k_\nu-\frac{k\cdot q}{q^2}q_\nu\right) \right\}\,,
\eeq
 so that the contribution from the dilaton is
 \beq
 k\frac{dN_\phi}{d^3\vec{k}}
 &=& \frac{e^2}{\pi^2 N_c^2Q^4} \frac{3}{8}2\kappa^2 \left(\frac{4}{3}\frac{1}{g_{YM}^2}\right)^2
\frac{\pi }{4k^4}\left(1-\frac{4k}{Q}\right)
  (p_\mu p'_\nu +p_\nu p'_\mu -\eta_{\mu\nu}p\cdot p')
  \nonumber \\
  && \times \left\{ (q\cdot k)^2 \left(\eta_{\mu\nu}-\frac{q_\mu q_\nu}{q^2} \right)
+q^2 \left(k_\mu-\frac{k\cdot q}{q^2}q_\mu \right)\left(k_\nu-\frac{k\cdot q}{q^2}q_\nu\right) \right\}
\nonumber \\
&=&\frac{\alpha_{em}}{16\pi^2 k^2}\left(\frac{1}{6} - \frac{2k}{3Q}\right)(1+\cos^2\theta)\,. \label{scalar}
\eeq

\subsection{The inclusive photon cross section}

The sum of (\ref{g}), (\ref{boson}) and (\ref{scalar}) is
\beq
k\frac{dN}{d^3\vec{k}}&=&k\frac{dN_G}{d^3\vec{k}}+k\frac{dN_A}{d^3\vec{k}} +k\frac{dN_\phi}{d^3\vec{k}}  \nonumber \\
&=&\frac{\alpha_{em}}{16\pi^2k^2}\Biggl\{ \left(\frac{7}{12}+\frac{1}{4}+\frac{1}{6} +\left(-2+0-\frac{2}{3}\right)\frac{k}{Q} \right)(1+\cos^2\theta) \nonumber \\
&& \qquad \qquad \qquad +\left(\frac{1}{2}+\frac{1}{2}+0 +\left(-1-1+0\right)\frac{k}{Q}\right)(1-\cos^2\theta) \Biggr\}
\nonumber \\
&=&\frac{\alpha_{em}}{8\pi^2k^2}\left(1-\frac{k}{3Q}(7+\cos^2 \theta)\right)
\nonumber \\
&\approx & \frac{\alpha_{em}}{8\pi^2k^2}\,. \label{main}
\eeq
Remarkably, the angular dependence has dropped out in the leading term $dN/dk\sim 1/k$, so the photon distribution in the $k\to 0$ limit  is spherical.

\section{Discussions}
One can take the view that the spherical distribution (\ref{main}) is quite reasonable because in this theory the integrated energy distribution in the final state of $e^+e^-$ annihilation is known to be exactly spherical  \cite{Hofman:2008ar}. [See, also, \cite{Lin:2007fa,Hatta:2008tx}.]   Nevertheless, the result is still nontrivial, even striking given that we have added dozens of terms from all the tensor components of the graviton, the gauge boson and the dilaton exchanges. None of the individual contributions are spherical, yet they  miraculously sum up with just the right proportions to generate a spherical distribution. To better appreciate the non-triviality, we note that the result of \cite{Hofman:2008ar} can be  entirely understood as being mediated by the graviton \cite{Hatta:2008st}.  This is not the case here--the gauge bosons and the dilaton exchanges are as important as the graviton contribution. Moreover, as (\ref{main}) also shows, the next--to--leading contribution is {\it not} spherical, so the sphericity of the leading term does not seem to follow immediately  from some symmetry argument.
Interestingly, the same thing happens in the weak (actually, zero) coupling calculation in Appendix A, namely, the distribution is not spherical in general, but it is so for the leading term (\ref{part}). Why only the leading term is spherical both at weak and strong coupling is puzzling  and  deserves further study.

Note that the collinear singularity in the weak coupling result (\ref{part}) has disappeared in the strong coupling result (\ref{main}) which is perfectly finite.  In fact, the absence of the collinear singularity  at strong coupling has been recurrently observed in the literature \cite{Strassler:2008bv,Hofman:2008ar,Hatta:2008tx,Hatta:2008tn,Hatta:2008st,Csaki:2008dt}. Our analysis provides another explicit confirmation of this phenomenon.

\section{Conclusions}
In this paper, we have demonstrated  a novel mechanism of soft photon production that is genuinely non-perturbative but fully under analytical control.  Since these  photons  follow the  $\sim 1/k$ distribution which is indistinguishable from  the usual Bremsstrahlung,   they could be the origin of the `anomalous' photons observed in the previous experiments. Of course,  the distribution in QCD is not spherical---the data show that the photon excess is primarily seen in the forward (low--$p_T$) region. However, the spherical distribution is most likely an artifact of ${\mathcal N}=4$ SYM, and will be lost under any attempt to deform the supergravity description in the spirit of AdS/QCD.  On the other hand,  our analysis suggests that the $1/k$ dependence presumably survives even when the sphericity is lost. Therefore, it would be very interesting to carry out  similar calculations in AdS/QCD models. We anticipate that this is not going to be an easy task, but requires many inputs from the  details of the collision before any comparison with the real data is possible. Nevertheless, given that the soft photon problem shows little sign of being resolved after more than 20 years since its  discovery, approaches based on gauge/string duality are certainly worth a try.


\section*{Acknowledgments}
 We are indebted  to Yuri Dokshitzer for drawing our attention to the soft photon problem.
 We thank Jian-Wei Qiu and Yuji Satoh for helpful conversations.
 This work is supported by Special Coordination Funds for Promoting Science and Technology of the Ministry of Education, the Japanese Government.
\appendix

\section{Photon production in weakly coupled ${\mathcal N}=4$ SYM}

In this Appendix, we calculate the inclusive cross section of photons
 \beq
  \label{formula} k\frac{d\sigma}{d^3\vec{k}}=\frac{4\pi \alpha_{em}^3}{Q^4}\left(W_T(1+\cos^2 \theta)+W_L (1-\cos^2\theta)\right)\,.
 \eeq
in ${\mathcal N}=4$ SYM at zero coupling (`parton model') using the ${\mathcal R}$--current (\ref{use}).
The fermionic terms in (\ref{use}) may be combined into a single Dirac fermion. Thus the calculation is the same as in QCD, except that the color factor $N_c$ is replaced by $N_c^2$. The result is known for a long time \cite{Walsh:1973mz}
\beq
W_{T}^{fermion}&=&\frac{N_c^2}{4}\frac{1}{4\pi^2}\left\{\frac{1+(1-x)^2}{x^2} \ln \frac{Q^2(1-x)}{m^2} -1 \right\}\,, \nonumber \\
W_L^{fermion}&=&\frac{N_c^2}{4}\frac{1}{4\pi^2}\frac{4(1-x)}{x^2}\,, \label{ff}
\eeq
 where $x\equiv 2k/Q$ is the Feynman variable and we have included the charge squared $\left(\frac{1}{2}\right)^2=\frac{1}{4}$ (see (\ref{use})). The parton mass $m$ is strictly zero in this theory, but we have included it in order to make the logarithm finite. The divergence in the limit $m\to 0$ is of course due to the collinear singularity, and one recognizes the usual splitting function in its prefactor. Note that the longitudinal structure function is nonzero even in the parton model.

Now consider the bosonic contribution $\gamma^*\to \phi\phi^\dagger \gamma$. In addition to the diagrams similar to the fermionic case, there is an extra diagram which involves a four--point contact interaction. After straightforward calculations, one gets  \beq
W_T^{boson}&=&2\frac{N_c^2}{4}\frac{1}{8\pi^2}\frac{1+(1-x)^2}{x^2}\,, \nonumber \\
W_L^{boson}&=&2\frac{N_c^2}{4}\frac{1}{4\pi^2} \left\{ \frac{1-x}{x^2}\ln \frac{Q^2(1-x)}{m^2}-3\frac{1-x}{x^2}\right\}\,,
\label{bb}
\eeq
 where the factor of 2 in front is because there are two complex scalars $\phi_{1,2}$. Contrary to the fermionic case, the logarithm appears in the longitudinal structure function. Summing (\ref{ff}) and (\ref{bb}) and substituting the result into (\ref{formula}), one finds that the distribution is not spherical in general. However, the leading logarithmic term in the soft limit $x\to 0$ is  spherical
 \beq
  k\frac{d\sigma}{d^3\vec{k}}\approx \frac{4\pi \alpha_{em}^3}{Q^4}\frac{N_c^2}{4\pi^2} \frac{1}{x^2}\ln \frac{Q^2}{m^2}\,.
  \eeq
 Dividing by the total cross section (\ref{tot}), one finds
 \beq
 k\frac{dN}{d^3\vec{k}} =\frac{\alpha_{em}}{2\pi^2 k^2}\ln \frac{Q^2}{m^2}\,, \label{part}
 \eeq
 which may be compared with the result at strong coupling (\ref{main}).

\section{Integral formulae}

Here we list the relevant integrals that appear in the intermediate calculations.
It is assumed that $\omega<Q$, and $Q$ has an infinitesimally small positive imaginary part.
\beq
 \frac{\pi Q}{2}\int_0^\infty dz\,  z^2 J_2(\omega z)H_1^{(1)}(Qz)
=\frac{-2i\omega^2}{(Q^2-\omega^2)^2}\,.    \label{a1}
\eeq
\beq
\frac{\pi Q}{2}\int_0^\infty dz\,  J_2(\omega z)H_1^{(1)}(Qz)
=-\frac{i}{2}
\left(1+\frac{Q^2}{\omega^2}\ln \left(1-\frac{\omega^2}{Q^2}\right)\right)\,.    \label{a2}
\eeq
\beq
\frac{\pi Q}{2}\int_0^\infty dz\, z J_1(\omega z)H_1^{(1)}(Qz)
=\frac{i\omega}{Q^2-\omega^2}\,.  \label{a3}
\eeq
\beq
\frac{\pi Q}{2}\int_0^\infty dz\, z^3 J_1(\omega z)H_1^{(1)}(Qz)
=\frac{-8i\omega Q^2}{(Q^2-\omega^2)^3}\,.  \label{a4}
\eeq
\beq
\frac{\pi Q}{2}\int_0^\infty dz\, z J_2(\omega z)H_0^{(1)}(Qz)
 =\frac{-i\omega^2}{Q}\left(\frac{1}{Q^2-\omega^2}+
\frac{\omega^2+Q^2\ln\left(1-\frac{\omega^2}{Q^2}\right)}{\omega^4}\right)\,.   \label{a5}
\eeq
\beq
\frac{\pi Q}{2}\int_0^\infty dz\, z^3 J_2(\omega z)H_0^{(1)}(Qz)
=\frac{8i\omega^2Q}{(Q^2-\omega^2)^3}\,.    \label{a6}
\eeq
\beq
\frac{\pi Q}{2}\int_0^\infty dz\, z^2 J_1(\omega z)H_0^{(1)}(Qz)
=\frac{-2i\omega Q}{(Q^2-\omega^2)^2}\,.   \label{a7}
\eeq

\providecommand{\href}[2]{#2}\begingroup\raggedright\endgroup


\begin{thebibliography}{10}

\bibitem{Landau:1953um}
L.~D. Landau and I.~Pomeranchuk, ``{Limits of applicability of the theory of
  bremsstrahlung electrons and pair production at high-energies},'' {\em Dokl.
  Akad. Nauk Ser. Fiz.} {\bf 92} (1953)
535--536.

\bibitem{Low:1958sn}
F.~E. Low, ``{Bremsstrahlung of very low-energy quanta in elementary particle
  collisions},'' {\em Phys. Rev.} {\bf 110} (1958)
974--977.

\bibitem{Chliapnikov:1984ed}
{\bf Brussels-CERN-Genoa-Mons-Nijmegen-Serpukhov} Collaboration, P.~V.
  Chliapnikov {\em et al.}, ``{Observation of direct soft photon production in K+p interactions at 70-GeV/c},'' {\em Phys. Lett.} {\bf B141} (1984)
276.

\bibitem{Botterweck:1991wf}
{\bf EHS-NA22} Collaboration, F.~Botterweck {\em et al.}, ``{Direct soft photon
  production in K+ p and pi+ p interactions at 250-GeV/c},'' {\em Z. Phys.}
  {\bf C51} (1991)
541--548.

\bibitem{Banerjee:1992ut}
{\bf SOPHIE/WA83} Collaboration, S.~Banerjee {\em et al.}, ``{Observation of
  direct soft photon production in pi- p interactions at 280-GeV/c},'' {\em
  Phys. Lett.} {\bf B305} (1993)
182--186.

\bibitem{Belogianni:1997rh}
{\bf WA91} Collaboration, A.~Belogianni {\em et al.}, ``{Confirmation of a soft
  photon signal in excess of QED expectations in pi- p interactions at
  280-GeV/c},'' {\em Phys. Lett.} {\bf B408} (1997) 487--492,
\href{http://arXiv.org/abs/hep-ex/9710006}{{\tt hep-ex/9710006}}.

\bibitem{Belogianni:2002ib}
  A.~Belogianni {\it et al.},
  ``{Further analysis of a direct soft photon excess in pi- p interactions at
  280-GeV/c},''
  {\em Phys.\ Lett.}\   {\bf B548} (2002) 122--128.

\bibitem{Belogianni:2002ic}
A.~Belogianni {\em et al.}, ``{Observation of a soft photon signal in excess of
  QED expectations in p p interactions},'' {\em Phys. Lett.} {\bf B548} (2002)
129--139.




\bibitem{Andersson:1988nk}
B.~Andersson, P.~Dahlqvist, and G.~Gustafson, ``{Soft photons in the Lund model},'' {\em Nucl. Phys.} {\bf B317} (1989)
635--646.

\bibitem{Shuryak:1989vn}
E.~V. Shuryak, ``{The `soft photon puzzle' and pion modification in hadronic matter},'' {\em Phys. Lett.} {\bf B231} (1989)
175--177.

\bibitem{Lichard:1990ye}
P.~Lichard and L.~Van~Hove, ``{The cold quark gluon plasma as a source of very
  soft photons in high energy collisions},'' {\em Phys. Lett.} {\bf B245}
  (1990)
605--608.

\bibitem{Botz:1994bg}
G.~W. Botz, P.~Haberl, and O.~Nachtmann, ``{Soft photons in hadron hadron
  collisions: Synchrotron radiation from the QCD vacuum?},'' {\em Z. Phys.}
  {\bf C67} (1995) 143--158,
\href{http://arXiv.org/abs/hep-ph/9410392}{{\tt hep-ph/9410392}}.

\bibitem{Pisut:1995vu}
  J.~Pisut, N.~Pisutova and B.~Tomasik,
  ``{Intermittent Behavior Of Bremsstrahlung Photons Produced In Hadronic
  Collisions At Very High-Energies},''
  {\em Phys.\ Lett.}\   {\bf B368} (1996) 179--186.



\bibitem{Aurenche:2006vj}
P.~Aurenche, M.~Fontannaz, J.-P. Guillet, E.~Pilon, and M.~Werlen, ``{A New
  critical study of photon production in hadronic collisions},'' {\em Phys.
  Rev.} {\bf D73} (2006) 094007,
\href{http://arXiv.org/abs/hep-ph/0602133}{{\tt hep-ph/0602133}}.

\bibitem{Abdallah:2005wn}
{\bf DELPHI} Collaboration, J.~Abdallah {\em et al.}, ``{Evidence for an excess
  of soft photons in hadronic decays of Z0},'' {\em Eur. Phys. J.} {\bf C47}
  (2006) 273--294,
\href{http://arXiv.org/abs/hep-ex/0604038}{{\tt hep-ex/0604038}}.

\bibitem{ko} V. Perepelitsa,    talk given at the XXXIX International Symposium on Multiparticle Dynamics, ``{perepelitza.ppt}"  available at  http://ismd2009.hep.by/ismd/talks/5/.

\bibitem{:2007zh}
{\bf DELPHI} Collaboration, J.~Abdallah {\em et al.}, ``{Observation of the
  Muon Inner Bremsstrahlung at LEP1},'' {\em Eur. Phys. J.} {\bf C57} (2008)
  499--514,
\href{http://arXiv.org/abs/0901.4488}{{\tt 0901.4488}}.

\bibitem{Wong:2010gf}
C.-Y. Wong, ``{QED2 Photons Associated with QCD String Fragmentation},''
\href{http://arXiv.org/abs/1001.1691}{{\tt 1001.1691}}.

\bibitem{Aharony:1999ti}
O.~Aharony, S.~S. Gubser, J.~M. Maldacena, H.~Ooguri, and Y.~Oz, ``{Large N
  field theories, string theory and gravity},'' {\em Phys. Rept.} {\bf 323}
  (2000) 183--386,
\href{http://arXiv.org/abs/hep-th/9905111}{{\tt hep-th/9905111}}.

\bibitem{Evans:2007sf}
N.~Evans and A.~Tedder, ``{A holographic model of hadronization},'' {\em Phys.
  Rev. Lett.} {\bf 100} (2008) 162003,
\href{http://arXiv.org/abs/0711.0300}{{\tt 0711.0300}}.

\bibitem{BallonBayona:2007rs}
C.~A. Ballon~Bayona, H.~Boschi-Filho, and N.~R.~F. Braga, ``{Deep inelastic
  structure functions from supergravity at small x},'' {\em JHEP} {\bf 10}
  (2008) 088,
\href{http://arXiv.org/abs/0712.3530}{{\tt 0712.3530}}.

\bibitem{Hatta:2008tx}
Y.~Hatta, E.~Iancu, and A.~H. Mueller, ``{Jet evolution in the N=4 SYM plasma
  at strong coupling},'' {\em JHEP} {\bf 05} (2008) 037,
\href{http://arXiv.org/abs/0803.2481}{{\tt 0803.2481}}.

\bibitem{Hatta:2008tn}
Y.~Hatta and T.~Matsuo, ``{Jet fragmentation and gauge/string duality},'' {\em
  Phys. Lett.} {\bf B670} (2008) 150--153,
\href{http://arXiv.org/abs/0804.4733}{{\tt 0804.4733}}.

\bibitem{Hatta:2008qx}
Y.~Hatta and T.~Matsuo, ``{Thermal hadron spectrum in $e^+e^-$ annihilation
  from gauge/string duality},'' {\em Phys. Rev. Lett.} {\bf 102} (2009) 062001,
\href{http://arXiv.org/abs/0807.0098}{{\tt 0807.0098}}.

\bibitem{Csaki:2008dt}
C.~Csaki, M.~Reece, and J.~Terning, ``{The AdS/QCD Correspondence: Still
  Undelivered},'' {\em JHEP} {\bf 05} (2009) 067,
\href{http://arXiv.org/abs/0811.3001}{{\tt 0811.3001}}.

\bibitem{Patino:2009uq}
L.~Patino and G.~Toledo, ``{Hadron production in electron-positron annihilation
  computed from the gauge gravity correspondence},'' {\em Phys. Rev.} {\bf D80}
  (2009) 126019,
\href{http://arXiv.org/abs/0901.4773}{{\tt 0901.4773}}.

\bibitem{Evans:2009py}
N.~Evans, J.~French, K.~Jensen, and E.~Threlfall, ``{Hadronization at the AdS
  wall},''
\href{http://arXiv.org/abs/0908.0407}{{\tt 0908.0407}}.

\bibitem{Keldysh:1964ud}
L.~V. Keldysh, ``{Diagram technique for nonequilibrium processes},'' {\em Zh.
  Eksp. Teor. Fiz.} {\bf 47} (1964)
1515--1527.

\bibitem{Balitsky:1990ck}
I.~I. Balitsky and V.~M. Braun, ``{The Nonlocal operator expansion for
  inclusive particle production in e+ e- annihilation},'' {\em Nucl. Phys.}
  {\bf B361} (1991)
93--140.

\bibitem{Freedman:1998tz}
D.~Z. Freedman, S.~D. Mathur, A.~Matusis, and L.~Rastelli, ``{Correlation
  functions in the CFT($d$)/AdS($d+1$) correspondence},'' {\em Nucl. Phys.}
  {\bf B546} (1999) 96--118,
\href{http://arXiv.org/abs/hep-th/9804058}{{\tt hep-th/9804058}}.

\bibitem{Bartels:2008zy}
J.~Bartels, A.~M. Mischler, and M.~Salvadore, ``{Four point function of
  R-currents in N=4 SYM in the Regge limit at weak coupling},'' {\em Phys.
  Rev.} {\bf D78} (2008) 016004,
\href{http://arXiv.org/abs/0803.1423}{{\tt 0803.1423}}.

\bibitem{Yoshida:2009dw}
Y.~Yoshida, ``{The Virtual Photon Structure Functions and AdS/QCD
  Correspondence},'' {\em Prog. Theor. Phys.} {\bf 123} (2010) 79--87,
\href{http://arXiv.org/abs/0902.1015}{{\tt 0902.1015}}.

\bibitem{Bartels:2009sc}
J.~Bartels, J.~Kotanski, A.~M. Mischler, and V.~Schomerus, ``{Regge limit of
  R-current correlators in AdS Supergravity},'' {\em Nucl. Phys.} {\bf B830}
  (2010) 153--178,
\href{http://arXiv.org/abs/0908.2301}{{\tt 0908.2301}}.

\bibitem{Cornalba:2009ax}
  L.~Cornalba, M.~S.~Costa and J.~Penedones,
  ``{Deep Inelastic Scattering in Conformal QCD},''
\href{http://arXiv.org/abs/0911.0043}{{\tt 0911.0043}}.


\bibitem{Cornalba:2010vk}
L.~Cornalba, M.~S. Costa, and J.~Penedones, ``{AdS black disk model for small-x
  DIS},''
\href{http://arXiv.org/abs/1001.1157}{{\tt 1001.1157}}.

\bibitem{CaronHuot:2006te}
S.~Caron-Huot, P.~Kovtun, G.~D. Moore, A.~Starinets, and L.~G. Yaffe, ``{Photon
  and dilepton production in supersymmetric Yang- Mills plasma},'' {\em JHEP}
  {\bf 12} (2006) 015,
\href{http://arXiv.org/abs/hep-th/0607237}{{\tt hep-th/0607237}}.

\bibitem{Gao:2009se}
J.-H. Gao and B.-W. Xiao, ``{Non-forward Compton scattering in AdS/CFT},''
\href{http://arXiv.org/abs/0912.4333}{{\tt 0912.4333}}.

\bibitem{Marquet:2010sf}
C.~Marquet, C.~Roiesnel, and S.~Wallon, ``{Virtual Compton Scattering off a
  Spinless Target in AdS/QCD},''
\href{http://arXiv.org/abs/1002.0566}{{\tt 1002.0566}}.

\bibitem{D'Hoker:1999jc}
E.~D'Hoker, D.~Z. Freedman, S.~D. Mathur, A.~Matusis, and L.~Rastelli,
  ``{Graviton and gauge boson propagators in AdS(d+1)},'' {\em Nucl. Phys.}
  {\bf B562} (1999) 330--352,
\href{http://arXiv.org/abs/hep-th/9902042}{{\tt hep-th/9902042}}.

\bibitem{D'Hoker:1999pj}
E.~D'Hoker, D.~Z. Freedman, S.~D. Mathur, A.~Matusis, and L.~Rastelli,
  ``{Graviton exchange and complete 4-point functions in the AdS/CFT
  correspondence},'' {\em Nucl. Phys.} {\bf B562} (1999) 353--394,
\href{http://arXiv.org/abs/hep-th/9903196}{{\tt hep-th/9903196}}.

\bibitem{Satoh:2002bc}
Y.~Satoh and J.~Troost, ``{On time-dependent AdS/CFT},'' {\em JHEP} {\bf 01}
  (2003) 027,
\href{http://arXiv.org/abs/hep-th/0212089}{{\tt hep-th/0212089}}.

\bibitem{Vermaseren:2000nd}
J.~A.~M. Vermaseren, ``{New features of FORM},''
\href{http://arXiv.org/abs/math-ph/0010025}{{\tt math-ph/0010025}}.

\bibitem{Hatta:2009ra}
Y.~Hatta, T.~Ueda, and B.-W. Xiao, ``{Polarized DIS in N=4 SYM: Where is spin
  at strong coupling?},'' {\em JHEP} {\bf 08} (2009) 007,
\href{http://arXiv.org/abs/0905.2493}{{\tt 0905.2493}}.

\bibitem{Hofman:2008ar}
D.~M. Hofman and J.~Maldacena, ``{Conformal collider physics: Energy and charge
  correlations},'' {\em JHEP} {\bf 05} (2008) 012,
\href{http://arXiv.org/abs/0803.1467}{{\tt 0803.1467}}.

\bibitem{Lin:2007fa}
S.~Lin and E.~Shuryak, ``{Toward the AdS/CFT Gravity Dual for High Energy
  Collisions: II. The Stress Tensor on the Boundary},'' {\em Phys. Rev.} {\bf
  D77} (2008) 085014,
\href{http://arXiv.org/abs/0711.0736}{{\tt 0711.0736}}.

\bibitem{Hatta:2008st}
Y.~Hatta, ``{Relating $e^+e^-$ annihilation to high energy scattering at weak
  and strong coupling},'' {\em JHEP} {\bf 11} (2008) 057,
\href{http://arXiv.org/abs/0810.0889}{{\tt 0810.0889}}.

\bibitem{Strassler:2008bv}
M.~J. Strassler, ``{Why Unparticle Models with Mass Gaps are Examples of Hidden
  Valleys},''
\href{http://arXiv.org/abs/0801.0629}{{\tt 0801.0629}}.

\bibitem{Walsh:1973mz}
T.~F. Walsh and P.~M. Zerwas, ``{Two photon processes in the parton model},''
  {\em Phys. Lett.} {\bf B44} (1973)
195--198.

\end{thebibliography}

\end{document}